# Dark Energy, Inertia and Mach's Principle

C Sivaram and Kenath Arun

Indian Institute of Astrophysics, Bangalore

**Abstract:** Mach's Principle says that a particle's inertia is due to some interaction of that particle with all the other masses in the universe. Here we explore the possibility of the gravitational interaction energy of the background quantum vacuum energy playing the role of a global Higg's field (described by a varying cosmological constant) entirely contributing to the local inertial masses of particles in the spirit of Mach's principle.

Mach's Principle implies that the local standards of non-acceleration are determined by some average of the motions of all the masses in the universe. As a result, it even implies interactions between inertia and electromagnetism. [1]

In the case of electromagnetic field, as is well known, an accelerated moving charge exerts a different electromagnetic force than a charge at rest (i.e. Coulomb field). This electromagnetic field, of a moving charge, falls of as $1/r$ rather than $1/r^2$ ! [2]

Thus we have for the force, for charges $q_1$ and $q_2$ separated by $r$:

$$F_{em} = \frac{kq_1q_2}{r^2} + \frac{kq_1q_2}{rc^2}a \qquad \ldots (1)$$

For large $r$, the $1/r$ term dominates. So we have:

$$F_{em} \approx \frac{kq_1q_2}{rc^2}a \qquad \ldots (2)$$

Now in the case of gravity similar force has been proposed (that is Einstein-Sciama force) for mass $m_1$ and $m_2$ separated by $r$: [3]

$$F_{Grav} = \frac{Gm_1m_2}{r^2} + \frac{Gm_1m_2}{rc^2}a \qquad \ldots (3)$$

And again for large r, the $1/r$ term dominates. That is:

$$F_{Grav} \approx \frac{Gm_1 m_2}{rc^2} a \qquad \ldots (4)$$

So if at all gravitational forces contribute to local inertia, this term dominates. So we do not consider the static $1/r^2$ term but only the $1/r$ term while considering the Machian effects of distant masses.

Thus we can write for the cumulative effect of all distant masses $m_i$ on the local mass $m$,

$$F_{Grav} = \sum_i \frac{Gm_i}{rc^2} ma \qquad \ldots (5)$$

which is summed over all the masses $m_i$

We can write for $\sum m_i = \int_{Vol} \rho dV$  ... (6)

$\rho$ is the average density of matter in the universe

$a$ is the acceleration

$c$ is the velocity of light

From this we have:

$$F_{Grav} = \frac{G}{c^2}\left[\int_{Vol} \frac{\rho dV}{r}\right] ma = \frac{G}{c^2}\left[\int_0^{R_H} \frac{\rho}{r}(4\pi r^2 dr)\right] ma \qquad \ldots (7)$$

(The integration is carried out over the Hubble volume)

This gives:

$$F_{Grav} = \frac{2\pi G \rho R_H^2}{c^2} ma = \frac{2\pi G \rho}{H_0^2} ma \qquad \ldots (8)$$

Where $H_0^2 = \left(\frac{c}{R_H}\right)^2 = \frac{8\pi G \rho_C}{3}$  ... (9)

and $\rho_C$ is the critical density

Therefore we have:

$$F_{Grav} = 0.75 ma \qquad \ldots (10)$$

If the universe is vacuum dominated or dark energy dominated [4], the simplest case being that of a cosmological constant, $\Lambda$ term, for which we have:

$$\rho = \frac{\Lambda c^2}{8\pi G}; \quad H_0^2 = \frac{\Lambda c^2}{3} \qquad \ldots (11)$$

This when substituted in equation (8) gives:

$$F = \frac{2\pi G \left(\frac{\Lambda c^2}{8\pi G}\right)}{\frac{\Lambda c^2}{3}} ma = 0.75 ma \qquad \ldots (12)$$

This indicates that gravitating vacuum energy (DE) could contribute up to 75% of the inertial mass of particles.

The quantum vacuum energy density for the curved space (of constant curvature) is given by: [5, 6]

$$\rho_{Vac} = \frac{\hbar c}{4\pi} \Lambda \int_0^{k_{max}} k\, dk + \hbar c \Lambda^2 \int_0^{k_{max}} \frac{dk}{k} + \hbar c \Lambda^3 \int_0^{k_{max}} \frac{dk}{k^3} + \ldots \qquad \ldots (13)$$

Since $\Lambda \approx 10^{-56} cm^{-2}$, the terms with higher powers of $\Lambda$ are much smaller. They become significant only at Planck scale. Therefore we have: [7]

$$\rho_{Vac} = \frac{\hbar c}{4\pi} \Lambda \int_0^{k_{max}} k\, dk \qquad \ldots (14)$$

If $k_{max} = \left(\frac{c^3}{\hbar G}\right)^{1/2}$ is the Planck cut off wavenumber $\sim \frac{1}{L_{pl}}$

$$\rho_{Vac} = \frac{\hbar c}{8\pi} \Lambda \left(\frac{c^3}{\hbar G}\right) = \frac{\Lambda c^4}{8\pi G} \qquad \ldots (15)$$

Using this in the equation (7) we have

$$F = \frac{G}{c^2}\left[\frac{\hbar c}{4\pi} \Lambda \int_0^{k_{max}} k\, dk \Big/ c^2\right]\left[\int_0^{R_H} \frac{4\pi r^2 dr}{r}\right] ma \qquad \ldots (16)$$

This gives: $F = \left(\Lambda R_H^2\right) ma$ $\qquad \ldots (17)$

This implies that $\Lambda R_H^2 \sim 1$, in order that all the local inertia is accounted for by the total gravitational interaction energy of the background vacuum energy up to Hubble radius.

This very important relation $(\Lambda R_H^2 \sim 1)$ is consistent with the present observations, that is:

$$\Lambda = \frac{1}{R_H^2} \approx 10^{-56} cm^{-2} \qquad \ldots (18)$$

Where $R_H = cH_0 = 10^{28} cm$ ... (19)

From equation (11) and (18) we have the dependence of dark energy density on the $\Lambda$ term as $\rho \propto \Lambda$, and $\Lambda \sim R_H^{-2} \propto (1+z)^2$. Therefore we have:

$$\rho_\Lambda = \rho_{\Lambda_0}(1+z)^2 \qquad \ldots (20)$$

(where $\rho_{\Lambda_0}$ is the present value of dark energy density)

Where as the matter density goes as:

$$\rho_m = \rho_{m_0}(1+z)^3 \qquad \ldots (21)$$

(where $\rho_{m_0}$ is the present value of matter density)

At a particular redshift, say, $z = z_0$, both matter and DE would have been equally significant.

$$\rho_{\Lambda_0}(1+z_0)^2 = \rho_{m_0}(1+z_0)^3 \qquad \ldots (22)$$

$$(1+z_0) = \frac{\rho_{\Lambda_0}}{\rho_{m_0}} \approx 3 \qquad \ldots (23)$$

That is $z_0 \approx 2$ ... (24)

This is consistent with observations as it has been observed that $\Lambda$ is dominant even at about 8 billion years ago corresponding to $z_0 \approx 2$. [8]

If $\Lambda$ were not dependent on $R_H$ (and hence $z$), the redshift where both matter and DE were equally significant would be given by:

$$\rho_{\Lambda_0} = \rho_{m_0}(1+z)^3 \qquad \ldots (25)$$

$z_0 \approx 0.44$, which is not quite consistent with observation.

At primordial nucleosynthesis, that is at $z \approx 10^9$, the energy densities are given by: [9]

$$\rho_\Lambda \approx 10^{18} \rho_{\Lambda_0}$$
$$\rho_m \approx 10^{27} \rho_{m_0} \qquad \ldots (26)$$
$$\rho_R \approx 10^{36} \rho_{R_0}$$

Thus at nucleosynthesis, the dark energy density was several orders lower than radiation energy density. So it will not affect element abundance.

At recombination, that is $z \approx 10^3$ the energy densities are given by:

$$\rho_\Lambda \approx 10^6 \rho_{\Lambda_0}$$
$$\rho_m \approx 10^9 \rho_{m_0} \qquad \ldots (27)$$
$$\rho_R \approx 10^{12} \rho_{R_0}$$

So $\rho_m$ and $\rho_R$ at recombination are several orders higher than $\rho_\Lambda$. So it will not much affect CMBR.

Only at around $z \approx 2$, $\rho_\Lambda$ would dominate.

Yet another way of arriving at the result, $\Lambda \sim 1/R_H^2$ is to invoke the fact that according to general relativity, the maximum force is given by: [10]

$$F_{max} = \frac{GM^2}{R_{min}^2} \qquad \ldots (28)$$

Where $R_{min} \approx \frac{GM}{c^2} \qquad \ldots (29)$

(that is, an object cannot be localised to $R < R_{min}$)

This implies that:

$$F_{max} = \frac{c^4}{G} \qquad \ldots (30)$$

This is independent of mass!

Applying this result (which holds independent of scale) over the entire Hubble radius gives for the cosmic energy density the relation:

$$\varepsilon = \frac{c^4}{GA} = \frac{c^4}{G(4\pi R_H^2)} \qquad \ldots (31)$$

Where $A \approx 4\pi R_H^2$ is the area of the Hubble surface.

The usual energy density implied by the cosmological constant is also given by:

$$\varepsilon = \frac{\Lambda c^4}{4\pi G} \qquad \ldots (32)$$

Comparing equations (31) and (32) again implies:

$$\Lambda \sim \frac{1}{R_H^2} \qquad \ldots (33)$$

Perhaps the first papers in which a time varying cosmological constant (DE or quintessence) were considered, i.e. $\Lambda$ varying as $1/R_H^2$ with epoch was ref. [11, 12].

Also, a time dependent vacuum energy with, $\Lambda \propto t^{-2}$ was obtained from a theory of unification of gravity with other interactions, where it was shown that the quantum vacuum energy (due to a Higgs field) owing to spontaneous symmetry breaking of a generalised gauge theory, in the early universe, varies with epoch as $t^{-2}$.

An exact solution was given as: [13]

$$R = R_0 \cosh^{3/2} \sqrt{\Lambda} t \qquad \ldots (34)$$

And this paper, written in 1975 was perhaps the very first solution which gives an inflationary exponential expansion driven by a large cosmological vacuum energy in the early universe.

In all the early papers, $\Lambda$ was $\Lambda_{Pl}(\sim 10^{66} cm^{-2})$ at $10^{-43} s$ (i.e. Planck time) and dropped down as $t^{-2}$, that is:

$$\Lambda_{present} \left( \approx \Lambda_{Pl} \left( \frac{t_{Pl}}{t_H} \right)^2 \right) \approx 10^{-56} cm^{-2} \qquad \ldots (35)$$

($t_H \sim 10^{17} s$ is the Hubble time)

Which is just the value deduced for the dominant dark energy density at present. For a summary of these early papers on dark energy by one of the present authors see ref. [14]

In conclusion, it is conceivable that the gravitational interaction energy of the background quantum vacuum energy playing the role of a global Higg's field (described by a varying cosmological constant) entirely contributes to the local inertial masses of particles in the spirit of Mach's principle.